\documentclass{article}
\usepackage[utf8]{inputenc}
\usepackage[english]{babel}

\usepackage{amsmath}
\usepackage{graphicx}
\graphicspath{ {./Images/} }
\usepackage[colorlinks=true, allcolors=blue]{hyperref}
\usepackage{amssymb}
\usepackage{dsfont}
\usepackage{algorithm}
\usepackage{algpseudocode}
\usepackage{bm}
\usepackage{amsthm}
\usepackage{arxiv}
\usepackage{lipsum}
\usepackage{float}
\usepackage{cite}
\usepackage{subcaption}
\usepackage{float}
\usepackage{url}
\usepackage{setspace}

\newcommand{\1}{\mathds{1}}

\def\1{\mathds{1}}

\newlength{\leftstackrelawd}
\newlength{\leftstackrelbwd}
\def\leftstackrel#1#2{\settowidth{\leftstackrelawd}%
{${{}^{#1}}$}\settowidth{\leftstackrelbwd}{$#2$}%
\addtolength{\leftstackrelawd}{-\leftstackrelbwd}%
\leavevmode\ifthenelse{\lengthtest{\leftstackrelawd>0pt}}%
{\kern-.5\leftstackrelawd}{}\mathrel{\mathop{#2}\limits^{#1}}}

\theoremstyle{plain} %Text ist Kursiv
\newtheorem{satz}{Theorem}[section]
\newtheorem{lem}[satz]{Lemma}
\newtheorem{thm}[satz]{Theorem}

\theoremstyle{definition}

\newtheorem{definition}[satz]{Definition}

\title{Optimal monotone conditional error functions}

\author{Werner Brannath\thanks{Equal contribution.} \\
	    Competence Center for Clinical Trials Bremen \& Institute of Statistics \\
        Faculty 3 - Mathematics and Computer Science\\
	    University of Bremen\\
	      \texttt{brannath@uni-bremen.de} \\
    \AND
       Johanna zur Verth\footnotemark[1] \\
       Competence Center for Clinical Trials Bremen \\
       Faculty 3 - Mathematics and Computer Science\\
       University of Bremen\\
       \texttt{jzurverth@uni-bremen.de} \\
    \AND
        Morten Dreher \\
	    Competence Center for Clinical Trials Bremen\\
        Faculty 3 - Mathematics and Computer Science\\
	    University of Bremen\\
	      \texttt{mdreher@uni-bremen.de} \\
    \AND
       Martin Scharpenberg \\
       Competence Center for Clinical Trials Bremen\\
       Faculty 3 - Mathematics and Computer Science\\
       University of Bremen\\
       \texttt{mscharpenberg@uni-bremen.de} \\
}

\renewcommand{\headeright}{A Preprint - \today}
\renewcommand{\undertitle}{}
\renewcommand{\shorttitle}{}

\hypersetup{
pdftitle={Optimal monotone conditional error functions},
pdfsubject={q-bio.NC, q-bio.QM},
pdfauthor={Werner Brannath, Morten Dreher, Johanna zur Verth, Martin Scharpenberg},
pdfkeywords={adaptive design, average sample size, conditional power, flexible design, likelihood ratio, maximum likelihood, unblinded sample size re-calculation},
}

\begin{document}
\maketitle

\setstretch{1.1}

\begin{abstract}
This paper presents a general method that provides optimal monotone conditional error functions for confirmatory adaptive two-stage designs with conditional power based sample size recalculations. The presented method builds on a previously developed general theory for optimal adaptive two-stage designs where sample sizes are reassessed for a specific conditional power and the goal is to minimize the expected sample size. The previous theory can easily lead to a non-monotonous conditional error function, which is highly undesirable for logical reasons and, as we show, can harm type I error rate control for composite null hypotheses. We also show that type I error control is generally guaranteed with a conditional error function (CEF)
that is non-increasing in the first stage p-value. We present a method that extends the existing theory by introducing an intermediate monotonising steps that can easily be implemented and provides a non-increasing conditional error function. We show mathematically that the monotonising step provides the optimal non-increasing conditional error function. We illustrate the method with several examples using \texttt{optconerrf}, an \texttt{R} package implemented for this paper.
\end{abstract}

\setstretch{1.2}

\section{Introduction}
In the planning phase of a clinical trial, only limited information regarding the treatment effect may be available. Therefore, one might want to choose a trial design with a data-driven sample size reassessment at an interim analysis. One approach to implement sample size reassessment is to use the conditional error function principle introduced in \cite{Proschan1995}. The idea is to calculate the conditional type I error rate for the second stage based on the evidence from the first stage, such that the overall type I error rate is under control. Most commonly used two-stage designs can be represented by conditional error functions. 
This includes two-stage combination tests, like  Fisher's product test and the inverse normal method, 
as well as classical group sequential designs, see e.g.\ \cite{Mueller2001}, \cite{Posch1999} and \cite{Wassmer2025}. 

Brannath and Bauer \cite{Brannath04} presented a theory for optimal adaptive two-stage designs with and without early stopping. The theory provides the conditional error function which minimizes the expected sample size when reassessing sample sizes for a specific conditional power. While the theory developed in \cite{Brannath04} is quite general, it does often lead to conditional error functions that are not monotone in the first stage p-value. Conditional error functions that are non-increasing
in the first stage p-value (or non-decreasing in the first stage Z-score), are highly desirable for logical reasons, and are required for type I error rate control with composite null hypotheses. The latter will be illustrated by examples and mathematically verified in Section~\ref{sec:NoMCER}.  In this paper we extend the optimality theory of Brannath and Bauer \cite{Brannath04} to conditional error functions that are constrained to be non-increasing in the first stage p-value also in cases where the unconstrained optimal conditional error function is not. We show that the optimal monotone conditional error function can be obtained by an `internal' monotonising step, preserving the general form of the optimal conditional error function and the numerical calibration procedure for type I error rate control. We will extend the theory to account for additional constraints like a minimum and maximum second stage sample size.

The paper is organized as follows. We start in the next section with a brief recap of the optimal conditional error functions introduced in Brannath and Bauer \cite{Brannath04}. In Section~\ref{sec:NoMCER} we discuss the requirement of a non-increasing conditional error function for type I error rate with conservative p-values. In Section~\ref{sec:omCEF} we present the optimal non-increasing conditional error function and describe its construction. Section~\ref{sec:examples} presents several examples for non-monotone optimal and optimal non-increasing conditional error functions with a comparison to the widely used inverse normal method. In Section~\ref{sec:bounds} we consider the optimization under bounding constraints on the second stage sample size and conditional error function. The main part of the paper ends with a discussion and points to some related open research topics. The Proofs of theorems are presented in the appendix. 

\section{Optimal conditional error functions}

We consider a confirmatory two-stage adaptive trial design with a single null hypothesis $H_0$; see e.g.\ \cite{Wassmer2025}. In the planning phase of the trial, we need to specify the overall significance level $\alpha$, the futility boundary $\alpha_0$, the early rejection boundary $\alpha_1 < \alpha_0$, and the conditional error function (CEF) with $0 \leq A(p_1) \leq 1$, which is a function of the first stage p-value $p_1$. We also must pre-specify for the first stage the sample size $n_1$, or more general, the first stage information $I_1$, as well the test procedure for the first stage p-value $p_1$.

At the first stage, the trial is stopped early for efficacy if $p_1 \leq \alpha_1$ and for futility if $p_1 > \alpha_0$. If the trial is not stopped early, we compute $A(p_1)$ and fix the trial design for the second stage based on the data from the first stage. 
The null hypothesis is rejected after the second stage if the second stage p-value $p_2$ fulfills $p_2 \leq A(p_1)$. We define $A(p_1):=1$ for $p_1 \leq \alpha_1$ and $A(p_1):=0$ for $p_1>\alpha_0$. 

In order to maintain the overall type I error rate $\alpha$ the p-values $p_1$ and $p_2$ need to satisfy the p-clud property under $H_0$
\begin{align*}
    \mathbb{P}(p_2 \leq u | p_1)\leq u \text{ for all } 0 \leq u \leq 1 \text{ and all } 0 \leq p_1 \leq 1
\end{align*}
and the conditional error function needs to satisfy the level condition
\begin{align} \label{levelcond}
    \int_0^1 A(p_1) dp_1 = \alpha_1 + \int_{\alpha_1}^{\alpha_0} A(p_1) dp_1 = \alpha.
\end{align}
Note that the level condition assumes a uniformly distributed first stage p-value.

We choose the sample size of the second stage so that we can reach a specific conditional power $cp$. For the interim sample size recalculation we need to make an assumption about the relative effect size $\delta$. A common choice is to use $\tilde{\delta}=\max(\hat{\delta}, \delta_0)$ where $\hat{\delta}$ is the first stage estimator of the relative effect size and $\delta_0$ is a minimally relevant effect. To be more general, we only assume for our theoretical results that the function $\tilde{\delta}(p_1)$ is a positive and non-increasing function in $p_1$, which is a rather natural assumption. The second stage sample size can then be calculated as
\begin{align*}
    n_2 = d\cdot\bigr[(\Phi^{-1}(cp)-\Phi^{-1}\{A(p_1)\}\bigl]^2/\tilde{\delta}(p_1)^2,
\end{align*}
where $\Phi^{-1}$ is the standard normal quantile function and $d$ is a design specific constant, which equals $d=2$ for the two-sample Z-test with balanced sample size. Brannath and Bauer \cite{Brannath04} proposed to minimize the expected sample size of the second stage, which can be written as
\begin{align}\label{expsamp}
  \mathbb{E}[n_2] = d \int_{\alpha_1}^{\alpha_0}\frac{\nu\{A(p_1)\}}{\tilde{\delta}(p_1)^2}l(p_1)\, dp_1,  
\end{align}
where 
\begin{align*}
    \nu(u)=\{\Phi^{-1}(cp)-\Phi^{-1}(u)\}^2 \text{ for } 0 \leq u < cp
\end{align*}
and $l(p_1)$ is the density of the first stage p-value $p_1$ under which the average sample size shall be minimized. If we want to minimize under the fixed effect size $\Delta$, the density $l(p_1)$ is, according to \cite{Hung1997}, given by
\begin{align}
    l_\Delta(p_1) =e^{\Phi^{-1}(1-p_1)\sqrt{I_1}\Delta - I_1 \Delta^2/2},
    \label{likelihoodratiofixed}
\end{align}
where $I_1$ is the first stage information. 
%In case of the balanced two-sample $Z$-test with $n_1$ subjects per group and a common variance of $\sigma^2$, it holds $I_1 = \frac{1}{2} \cdot \frac{n_1}{\sigma^2}$. 

It has been illustrated in \cite{Brannath04} that minimizing the expected second stage sample size under a fixed parameter $\Delta$ can lead to rather inefficient designs when $\Delta$ is not correctly specified.
An alternative is to choose a Bayesian prior distribution for the parameter $\Delta$ and obtain the corresponding density of the first stage p-value. For a normal, exponential, or uniform prior distribution the corresponding functions $l(p_1)$ can be found in \cite{Wassmer2025}. Another possibility is to use the maximum likelihood estimate for $\max(\Delta,0)$, which leads to
\begin{align}
    l_{ML}(p_1)= e^{\max\{0, \Phi^{-1}(1-p_1)\}^2/2},
    \label{likelihoodratioML}
\end{align}
and  has the advantage that we do not need to choose a prior distribution.

Brannath and Bauer \cite{Brannath04} minimize the expected sample size (\ref{expsamp}) under the following constraints
\begin{align*}
    &\text{(C1)} \quad 0 \leq A(p_1) < cp \text{ for all } p \in [\alpha_1, \alpha_0]\\
    &\text{(C2)} \quad \text{The level condition (\ref{levelcond}) is fulfilled.}
\end{align*}
The first constraint (C1) ensures that the conditional type I error rate never exceeds the conditional power. The second constraint (C2) needs to be fulfilled to keep the overall type I error rate. They show that under the assumption $1-\Phi(2)\le cp \le \Phi(2)$ the possibly non-monotone optimal conditional error function that fulfills the constraints (C1) and (C2) has the form 
\begin{align}\label{eq:oce}
A_{opt,c}(p_1)= \psi\{-e^{c}/ Q(p_1)\},
\end{align}
where $Q(p_1)=l(p_1)/\tilde{\delta}(p_1)^2$ and $\psi$ is the inverse of the derivative $\nu'$ of the function $\nu$. For $1-\Phi(2)\le cp \le \Phi(2)$ the derivative $\nu'$ is increasing and therefore invertible. The constant $c$ needs to be chosen so that $A_{opt,c}$ fulfills the level condition \eqref{levelcond}.
Fortunately, the assumption $1-\Phi(2)\le cp \le \Phi(2)$ is not required for the derivation of the optimal conditional error function. In the next theorem we show that for 
$cp> \Phi(2)$ and for $cp<1-\Phi(2)$ the optimal conditional error function is given by (\ref{eq:oce}) with a specific generalization $\tilde{\psi}$ of $\psi$. The proof can be found in the appendix.
\begin{thm}\label{nocpconstr}
Let us assume $cp > \Phi(2)$ or $cp < 1-\Phi(2)$ and define
\begin{align*}
u_{1/2} := 1- \Phi \{-\Phi^{-1}(cp)/2 \pm \sqrt{\Phi^{-1}(cp)^2/4-1}\}
\end{align*}
with $u_1 < u_2$.
The optimal conditional error function is given by (\ref{eq:oce}) with $\psi$ replaced by  
\begin{align}
    \tilde{\psi}(x)=
    \begin{cases}
    \psi_l &\text{for } x < \nu'(u_2) \text{ or } x < \frac{\nu(\psi_u)-\nu(\psi_l)}{\psi_u-\psi_l} \\
    \psi_u &\text{for } x> \nu'(u_1) \text{ or } x \geq \frac{\nu(\psi_u)-\nu(\psi_l)}{\psi_u-\psi_l},\\
    \end{cases}
\end{align}
where $\psi_l=\psi_l(x)$ is the solution of the equation $\nu'(u) = x $ in the interval $]0, u_1]$ and 
$\psi_u=\psi_u(x)$ is the solution of the equation $\nu'(u) = x$ in the interval $[u_2, cp[$. The function $\tilde{\psi}(\cdot)$ is non-decreasing. 
\end{thm}

\section{Requirement of non-increasing conditional error functions}\label{sec:NoMCER}
A non-increasing conditional error function is highly desirable for logical reasons: Rejection of $H_0$ at the second stage should become more difficult with less evidence against $H_0$ from the first stage. In other words, the second stage level 
$A(p_1)$ should never increase with decreasing first stage evidence, i.e.\ should nowhere be increasing in $p_1$. 
%A conditional error function that is increasing in the first stage p-value could lead to a situation where we could have rejected the null hypothesis at the second stage more easily when the p-value of the first stage had been larger. 
A partially increasing conditional error function not only leads to logical inconsistencies, but also does not necessarily keep the overall type I error rate. Consider the composite null hypothesis $H_0: \delta \leq 0$. The first stage p-value is uniformly distributed for $\delta=0$ and usually strictly conservative for all other parameter values in the null hypothesis in the following sense:
\begin{definition}\label{conspvalue}
    A p-value is said to be `strictly conservative' if the distribution function $F_c(x)$  of the p-value satisfies the following:
%    \begin{enumerate}
%        \item  
$$ F_c(x) \leq x \text{ for all } 0\le x\le 1 \quad\text{and}\quad F_c(x) < x \text{ for at least one } 0 < x < 1 \,. 
%        \item 
$$
%    \end{enumerate}
\end{definition}
For a trial design without an interim analysis, a conservative p-value does not fully exploit the significance level and, in particular, keeps the overall type I error rate. Theorem \ref{inflation} below (proved in the appendix) shows that this is not necessarily the case for a two-stage design with a conditional error function that is increasing on some interval, even though it satisfies level condition (\ref{levelcond}).
\begin{thm}\label{inflation}
If $A(p_1)$ is a conditional error function that satisfies level condition (\ref{levelcond}) and is increasing on a non-empty interval
$]d_1, d_2[ \subseteq ]\alpha_1, \alpha_0]$, then there exists a density $f_c(p_1)$ of a conservative first-stage p-value $p_1$ that leads to a type I error inflation with $A(p_1)$.
\end{thm}
We illustrate the type I error rate inflation with a concrete example. For simplicity, we consider a design without early rejection and futility boundaries. We test the null hypothesis $H_0: \delta \leq 0$ and assume a first stage test statistic $Z_1\sim N(\sqrt{10} \delta, 1)$ with corresponding first stage p-value $p_1 = 1- \Phi(Z_1)$. This corresponds to a first stage design with two balanced treatment groups, each with sample size $n_1=20$. Obviously, the first-stage p-value is conservative in sense of Definition~\ref{conspvalue} for all $\delta <0$ (with $F_c(x)<x$ for all $0< x <1$). The actual type I error rate can be calculated numerically by
\begin{align*}
 \alpha^* = \int_{0}^1 f(p_1) A(p_1) dp_1,
\end{align*}
with $f(p_1)=l_\delta(p_1)$ the density of the first-stage p-value, where $l_\delta(p_1)$ as in \eqref{likelihoodratiofixed} with $\Delta$ replaced by $\delta$.
We target the overall significance level of $\alpha = 0.05$ (for $\delta=0$) and determine the corresponding optimal conditional error function \eqref{eq:oce} using the maximum likelihood ratio \eqref{likelihoodratioML} with a sample size reassessment for conditional power 0.8 at the interim estimate with lower cut-off $\delta_0 = 0.125$. The resulting conditional error function is not monotone, more precisely, increases on some interval of $[0,1]$. We calculate the type I error rate for the mean shifts $\delta= -0.25$ and $\delta = -0.5$ for this optimal CEF and, for a comparison, also for the inverse normal conditional error functions with equal weights ($w_1 =w_2 =\sqrt{1/2}$) and unequal weights $w_1 =\sqrt{1/3}$, $w_2 =\sqrt{2/3}$ which is, in both cases, non-increasing for all $p_1\in [0,1]$. The results are presented in Table~\ref{tab:simtypeI}.

\begin{table}%[H]
\renewcommand{\arraystretch}{1.5}
\begin{center}
\begin{tabular}{c|ccc} 
Mean shift ($\delta$) &
 \parbox{3cm}{\centering Inv. normal \\($w_1=w_2=\sqrt{1/2}$)} & 
 \parbox{4cm}{\centering Inv. normal \\($w_1= \sqrt{1/3}$, $w_2=\sqrt{2/3}$)} &
 \parbox{3cm}{\centering Optimal \\(Max. likel.)}\\ 
  \hline
 0.0 & 0.050 & 0.050 & 0.050\\ 
 -0.25 & 0.014  & 0.018 & 0.059\\
 -0.5 & 0.003 & 0.005 & 0.062\\ 
\end{tabular}
\end{center}
\caption{Approximations of the type I error rate for $\delta=0$, $\delta= -0.25$ and $\delta = -0.5$ for the inverse normal conditional error function with equal weights ($w_1=w_2=\sqrt{1/2}$), the inverse normal conditional error function with the weights $w_1 = \sqrt{1/3}$, $w_2 =\sqrt{2/3}$ and the optimal conditional error function (maximum LR, $\delta_0= 0.125$, $cp = 0.8$) }
\label{tab:simtypeI}
\end{table}

One can see from Table \ref{tab:simtypeI} that both inverse normal methods keep the type I error rate in all scenarios with a type I error rate falling far below the overall significance level for the $\delta<0$. The optimal conditional error function only keeps the type I error rate for $\delta=0$ and clearly exceeds the type I error rate for $\delta = -0.25$ and $\delta = -0.5$. The following theorem recalls the well-known fact that the type I error is under control also for conservative first-stage p-values whenever the conditional error function is non-increasing in $p_1$ on the whole continuation region. For completeness, the proof is given in the appendix. 
\begin{thm}\label{typeIerror}
With a non-increasing conditional error function $A(p_1)$ satisfying level condition \eqref{levelcond} and a conservative first stage p-value $p_1$ with density $f_c(p_1)$, we obtain $\int_0^1 A(p_1) f_c(p_1)\,dp_1 \leq \alpha$.
\end{thm}

Given the presented issue with non-monotone conditional error functions, we focus in the next section on the construction of non-increasing  optimal conditional error functions, i.e.\ non-increasing conditional error functions that minimize the expected sample size under a given density or (the maximum likelihood ratio) for the first stage p-value $p_1$, when reassessing samples for a specific target conditional power.

\section{Optimal monotone conditional error functions}\label{sec:omCEF}

In addition to constraints (C1) and (C2), we aim now to satisfy the constraint:
\begin{align*}
    \text{(C3)} \quad A(p_1) \text{ is non-increasing in } p_1 \text{ for all }p_1\in ]\alpha_1,\alpha_0].
\end{align*}
If $Q(p_1)$ is non-increasing in $p_1$, then
$A_{opt,c}= \psi\{-e^{c}/ Q(p_1)\}$ is non-increasing as
well (since $\psi\{\cdot\}$ is increasing), and is optimal
according to Theorem~4.1 in \cite{Brannath04}. If $Q(p_1)$ is increasing on some
intervals, $A_{opt,c}(p_1)$ is increasing at the same
intervals, and hence does not fulfill (C3) anymore. We will derive the optimal
{\em non-increasing} conditional error function for the case where
$Q(p_1)$ is increasing on a finite number of disjoint subintervals
$D_k=\,]d_{l k},d_{u k}]$ ($k=1,\ldots, K$) of the continuation
region $]\alpha_1,\alpha_0]$ and is non-increasing outside these intervals. To this aim we modify
$Q(p_1)$ to a suitable {\em non-increasing} function
$\tilde{Q}(p_1)$ which is constant on each $D_k$, and then show
that the optimal {\em non-increasing} conditional error function
is given by
$\tilde{A}_{opt, c_{\alpha}}(p_1)=\psi\{-e^{c_\alpha}/\tilde{Q}(p_1)\}$.

\subsection{Monotonising the function $Q(p_1)$}\label{Qmono}
We construct $\tilde{Q}(\cdot)$ by the following stepwise inductive procedure.
We first define
\begin{align}\label{eq:tildeQ}
\tilde{Q}^{(1)}_{q}(p_1):=
\left\{\begin{array}{ll}
  \max\{q,Q(p_1)\}  &\text{for }  p_1\le d_{l 1}  \\
   q                &\text{for }  d_{l 1}< p_1\le d_{u 1} \\
   \min\{q,Q(p_1)\} &\text{for }  d_{u 1}< p_1\le d_{l 2}\\
   Q(p_1) &\text{for }  p_1> d_{l 2}
\end{array}\right.,
\end{align}
where $q$ is a positive number, and $d_{l 2}:=\alpha_0$ if $K=1$.
Then we choose the unique positive number $q_1$ such that
$\int_{\alpha_1}^{d_{l 2}}\,\tilde{Q}_{q_1}^{(1)}(p_1)\,dp_1=
\int_{\alpha_1}^{d_{l 2}}\,Q(p_1)\,d p_1$. Such a choice %of $m_1$
is always possible, since the integral on the left side increases
continuously from 
$\int_{\alpha_{1}}^{d_{l1}} Q(p_1)dp_1< \int_{\alpha_1}^{d_{l2}} Q(p_1)dp_1$
%$\int_{d_{u1}}^{d_{l2}} Q(p_1)dp_1< \int_{\alpha_1}^{d_{l2}} Q(p_1)dp_1$
to $\infty$ if $q_1$ increases from 
0 to
$\infty$. By definition, $\tilde{Q}^{(1)}_{q_1}(p_1)$ is
non-increasing on $]\alpha_1,d_{l 2}]$. In the case $K=1$ we
are finished, since $d_{l 2}=\alpha_0$. If, in particular,
$Q(p_1)$ is increasing on the whole continuation region
($D_1=\,]\alpha_1,\alpha_0]$), then
$\tilde{Q}^{(1)}_{q_1}(p_1)$ is identical to the constant
$q_1=\int_{\alpha_1}^{\alpha_0}\,Q(p_1)\,dp_1/(\alpha_0-\alpha_1)$. If $K\ge
2$ then $\tilde{Q}^{(1)}(p_1)$ is still decreasing on $D_k$ for
$k\ge 2$, and we continue inductively defining for $k=2, \ldots,
K$ 
\begin{equation}\label{tm}
\tilde{Q}^{(k)}_{q_k}(p_1):=
\left\{\begin{array}{ll}
  \max\{q_k,\tilde{Q}^{(k-1)}_{q_{k-1}}(p_1)\}  &\text{for }  p_1\le d_{l k}  \\
   q_k                &\text{for }  d_{l k} < p_1\le d_{u k} \\
   \min\{q_k,Q(p_1)\} &\text{for }  d_{u k} < p_1\le d_{l k+1}\\
   Q(p_1) &\text{for }  p_1 > d_{l k+1}
\end{array}\right.\qquad (\mbox{where } d_{l K+1}:=\alpha_1)
\end{equation}
and choose $q_k$ such that $\int_{\alpha_1}^{d_{l k+1}}\tilde{Q}^{(k)}_{q_k}(p_1)\,d p_1=
\int_{\alpha_1}^{d_{l k+1}}\, Q(p_1)\,d p_1$. For a visualization of monotonising the function $Q(p_1)$, we refer to figure \ref{conderfun}\textbf{A}.
\subsection{Optimal non-increasing conditional error function}
%Next we show that the optimal {\em non-increasing} conditional error function
%is given by $\tilde{A}_{opt,c_{\alpha}}(p_1)=\psi\{-e^{c_\alpha}/\tilde{Q}(p_1)\}$. 
The following theorem states that the optimal non-increasing conditional error function is given by $\tilde{A}_{opt,c_{\alpha}}(p_1)=\psi\{-e^{c_\alpha}/\tilde{Q}(p_1)\}$. The proof of the theorem can be found in the appendix.
\begin{thm}\label{opti} Let $\psi(\cdot)$ be as in (\ref{eq:oce}) (respectively for  $cp> \Phi(2)$ or $cp<1-\Phi(2)$ as in Theorem  \ref{nocpconstr}), $Q(p_1)=l(p_1)/\tilde{\delta}(p_1)^2$,
and $\tilde{Q}(p_1)$ the non-increasing modification of $Q(p_1)$
as defined in \eqref{eq:tildeQ} and, if required, in \eqref{tm}. Let
$\tilde{A}_{opt, c_\alpha}(p_1)=\psi\{-e^{c_\alpha}/\tilde{Q}(p_1)\}$
with $c_\alpha$ such that $\tilde{A}_{opt, c_\alpha}(\cdot)$
satisfies the level condition (\ref{levelcond}). Then
$\tilde{A}_{opt, c_\alpha}(p_1)$ is non-increasing on
$]\alpha_1,\alpha_0]$, and for every other {\em
non-increasing} conditional error function $A(p_1)$ which
satisfies level condition (\ref{levelcond})
$$\int_{\alpha_1}^{\alpha_0}\,\nu\{\tilde{A}_{opt, c_\alpha}(p_1)\}\,Q(p_1)\,d p_1<
\int_{\alpha_1}^{\alpha_0}\,\nu\{A(p_1)\}\,Q(p_1)\,d p_1.
$$
This implies that \eqref{expsamp} is minimized by the non-increasing conditional error function $\tilde{A}_{opt, c_\alpha}$.
\end{thm}

The function $\tilde{A}_{opt, c_\alpha}(p_1)$ can be determined numerically. To determine $\tilde{Q}(p_1)$ we first
determine all intervals $]d_{l k},d_{u k}]$
(e.g.\ by numerical root finding), and
at each inductive step the constant $q_k$ (by numerical integration and root finding).
In the case where $\tilde{\delta}(p_1)$ is the observed treatment effect at the interim analysis (truncated from below by some constant $\delta_0>0$), we have always observed at most
two intervals of decrease, so that $\tilde{Q}(p_1)$ could easily be determined numerically. As indicated previously, the method and the optimal conditional error $\tilde{A}_{opt, c_\alpha}(p_1)$ are implemented in the \texttt{R}-package \texttt{optconerrf} \cite{R}, \cite{optconerrf}.

\section{Examples and comparisons}\label{sec:examples}
In the following, we show different examples for optimal monotone conditional error functions using our \texttt{R}-package \texttt{optconerrf}. An introduction to the usage of the package \texttt{optconerrf} can be found in the vignettes \cite{introoptconerrf} and \cite{furtheroptconerrf}. Furthermore, we compare the optimal monotone conditional error functions to potentially non-monotone ones and to other more standard approaches.

\subsection{Numerical example}\label{sec:example}
Consider planning an adaptive two-stage design with stage-wise p-values from one-sided, two-sample $Z$-tests. We aim on equal group sizes and assume a common variance of $\sigma^2=1$ in both treatment groups. The overall significance level is $\alpha = 0.05$ and we want to achieve an overall power of $0.8$ at the alternative $\delta = 0.2$. In case of a fixed-size-sample test, a sample size of $N = 2\{\Phi^{-1}(0.8)- \Phi^{-1}(0.05)\}^2/0.2^2 \approx 310$ per group is required. We plan an adaptive two-stage design with roughly a third of this sample size in the first stage ($n_1 = 104$). Furthermore, we want to incorporate the early rejection boundary $\alpha_1=0.001$ and a futility boundary $\alpha_0=0.5$. We aim for a conditional power of $cp = 0.8$ and use for the interim calculation of the second stage sample size the interim estimate $\tilde{\delta}=\max(\hat{\delta}_1,\delta_0)$ of the effect size with lower cut-off value $\delta_0=0.125$. We determine the conditional error function which minimizes the expected sample size for the effect size  $\Delta = 0.2$, i.e. assume density \eqref{likelihoodratiofixed} with $\Delta=0.2$ in \eqref{expsamp}.

\begin{figure}%[H]
\includegraphics[width=1\linewidth]{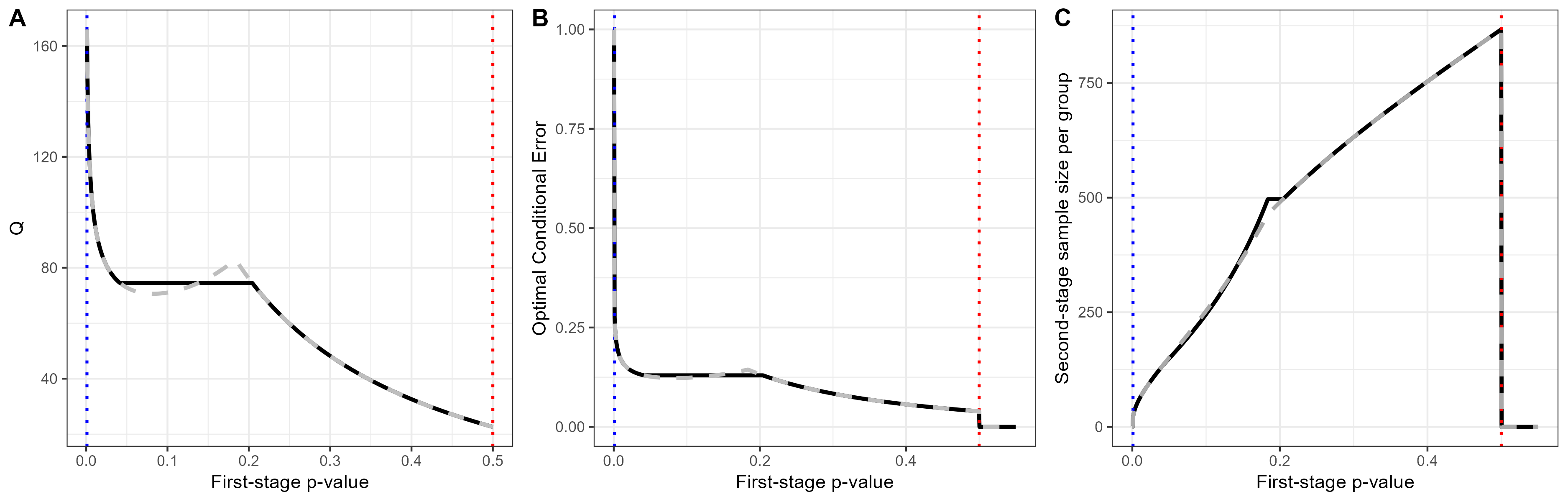}
\centering
\caption{\textbf{A:} Unconstrained Q (dashed line) and monotone Q (solid line) \textbf{B:} Unconstrained optimal CEF (dashed line) and optimal monotone CEF (solid line) for the example described in the text, \textbf{C:} Corresponding second stage sample sizes}
\label{conderfun}
\end{figure}

For the calculation with the \texttt{R}-package \texttt{optconerrf}, it is required to specify the first stage information $I_1$ instead of the group-wise first stage sample size $n_1$, which equals $I_1 = n_1/(2\sigma^2)$ in this example. Figure \ref{conderfun} \textbf{A} shows $Q(p_1)$ and the non-increasing modification of $Q(p_1)$ (see chapter \ref{Qmono}). Figure \ref{conderfun} \textbf{B} shows the resulting unconstrained optimal CEF (dashed line) and the optimal monotone CEF (solid line). The unconstrained optimal CEF is non-monotone on a single interval, and so the optimal one is constant on an interval, namely on $[0.04, 0.20]$ with the constant value $q = 74.56$ in \eqref{eq:tildeQ}. The level constant for the optimal monotone conditional error function is $c_\alpha = 7.24$. 

In figure \ref{conderfun} \textbf{C} the corresponding group-wise second stage sample size is plotted as a function of the first stage p-value for the optimal conditional error function (dashed line) and the optimal monotone conditional error function (solid line). The package provides only a plot of the second stage information. The plot of the second stage sample size was obtained with a minor modification of the implementation.

\begin{figure}%[H]
\includegraphics[width=.8\linewidth]{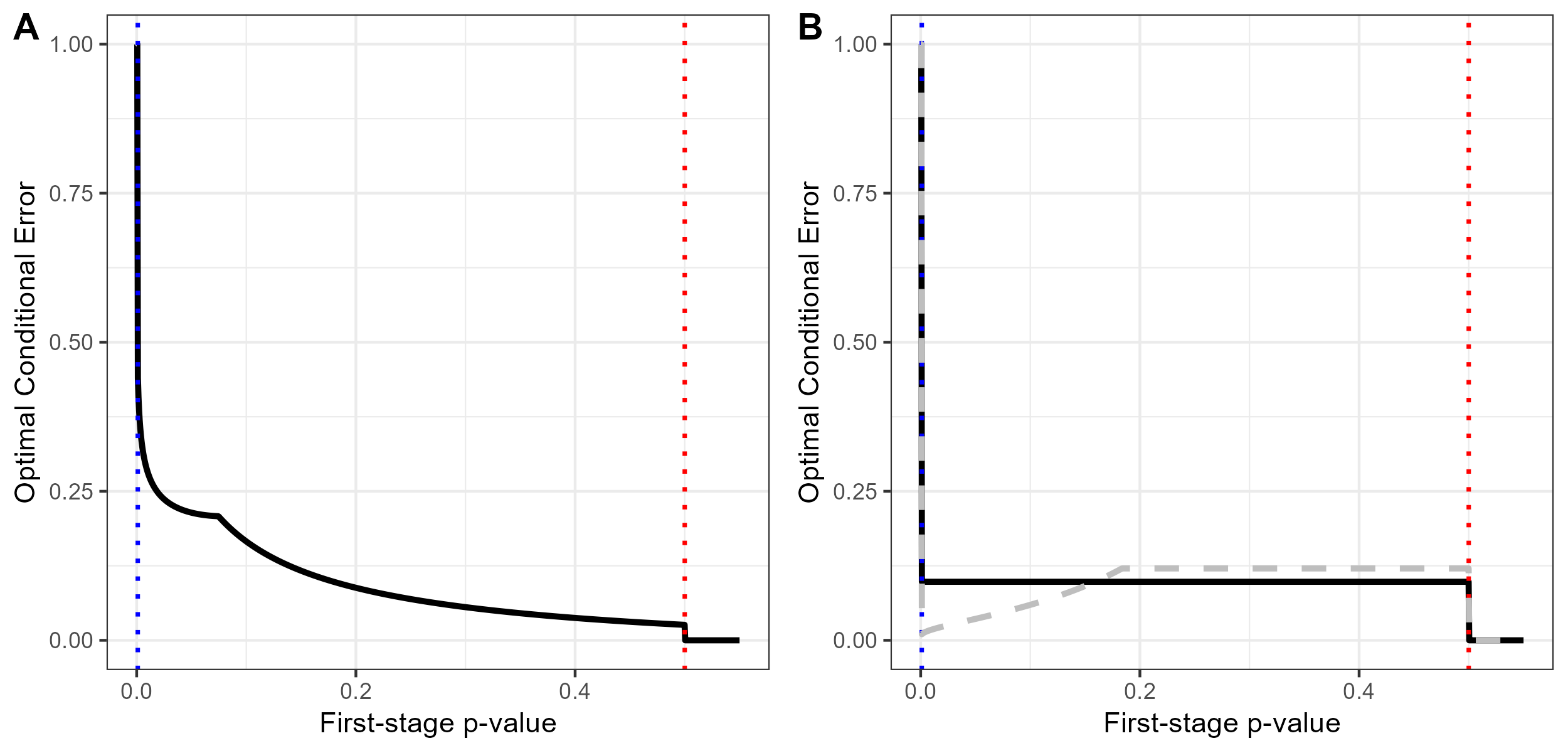}
\centering
\caption{\textbf{A:} Unconstrained optimal CEF for $\Delta = 0.2$ and $\delta_0 = 0.2$, \textbf{B:} Unconstrained optimal CEF (dashed line) and optimal monotone CEF (solid line) for $\Delta = 0$ and $\delta_0 = 0.125$.}
\label{conderfun_alt}
\end{figure}

The unconstrained optimal CEF becomes monotone when choosing a larger lower cut-off value $\delta_0$. In the above example, setting $\delta_0 = 0.2$ results in an unconstrained optimal CEF that is already monotone (see figure \ref{conderfun_alt} \textbf{A}). We also note that minimizing the expected second stage sample size for a value of $\Delta$ in the null hypothesis, we obtain a constant optimal monotone CEF on the whole continuation region $[\alpha_0, \alpha_1]$, because the unconstrained optimal CEF is increasing. Figure \ref{conderfun_alt} \textbf{B} shows the optimal conditional error function for the above example when optimizing under the null $\Delta = 0$.

\subsection{Comparisons of methods}\label{compmethods}

\begin{figure}%[H]
\includegraphics[width=.8\linewidth]{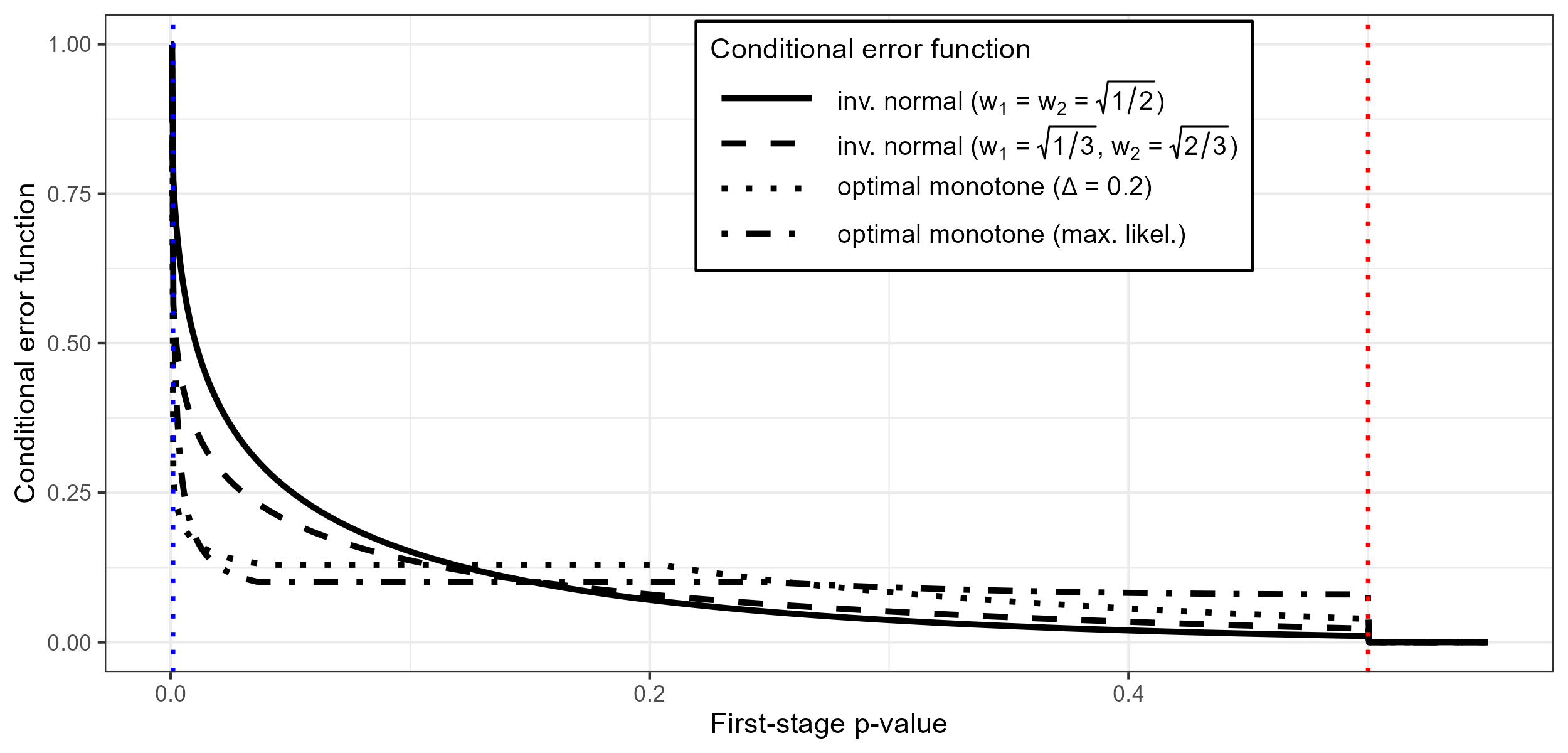}
\centering
\caption{CEF of the inverse normal method with equally weighted stages ($w_1=w_2= \sqrt{1/2}$) and unequal weights $w_1 = \sqrt{1/3}$, $w_2 = \sqrt{2/3}$, optimal monotone CEF for effect size $\Delta = 0.2$, optimal monotone CEF with maximum likelihood approach. The inverse normal method is currently not implemented in \texttt{optconerrf} and has been added here.}
\label{conderfun_different}
\end{figure}

Instead of using the fixed effect size $\Delta = 0.2$ for the density of the first stage p-value, it may be preferable to use the maximum likelihood approach which does not require any additional assumptions. In the following, we will compare these two approaches. In addition, we consider the inverse normal method with equally weighted stages ($w_1 = w_2 = \sqrt{1/2}$) and also with the weights $w_1 = \sqrt{1/3}$ and $w_2 = \sqrt{2/3}$. The choice of the latter (unequal) weights is motivated by our decision to use a third of the single-stage sample size for the first stage. Figure~\ref{conderfun_different} shows the four different conditional error functions. 

The maximum second stage sample size is largest for the inverse normal method with equally weighted stages ($1272$), smaller for the inverse normal method with the weights $w_1=\sqrt{1/3}$, $w_2=\sqrt{2/3}$ ($1010$), and even smaller with the optimal CEF for $\Delta = 0.2$ ($868$) and smallest for the maximum likelihood approach ($647$).

In table \ref{expectedsecondstagesamplesize} the expected second stage sample size for a true $\delta$ of $0$, $0.125$ and $0.2$ is displayed and in parentheses the corresponding overall power is given. The optimal monotone conditional error function that uses the fixed effect size $\Delta = 0.2$ for the density of the first stage p-value has the smallest expected second stage sample size when $\delta=0.2$  is true. The expected second stage sample size of the optimal monotone CEF with the maximum likelihood approach is the second smallest and not much larger than the minimal one, and it provides the smallest sample size for the other true $\delta$ values. This is in line with the numerical findings in  \cite{Brannath04}.

\begin{table}%[H]
\centering
\renewcommand{\arraystretch}{1.5}
\begin{tabular}{c|cccccc}
$\delta$ &
 \parbox{2cm}{\centering Inv. normal \\($w_1=w_2 \newline =\sqrt{1/2}$)} & 
 \parbox{2cm}{\centering Inv. normal \\($w_1= \sqrt{1/3}$, $w_2=\sqrt{2/3}$)} &
 \parbox{2cm}{\centering Optimal monotone \\($\Delta = 0$)}&
 \parbox{2cm}{\centering Optimal monotone \\($\Delta = 0.125$)}&
 \parbox{2cm}{\centering Optimal monotone \\($\Delta = 0.2$)}&
 \parbox{2cm}{\centering Optimal monotone \\(Max. likel.)}\\ 
  \hline
0 & 348.08 (0.05) & 308.77 (0.05)& 236.86 (0.05) & 242.22 (0.05) & 259.39 (0.05)& 245.24 (0.05)\\ 
0.125 & 375.18 (0.57) & 344.48 (0.56) & 297.07 (0.51)& 295.00 (0.51) & 301.04 (0.53) & 299.54 (0.53)\\ 
0.2 & 283.52 (0.78)& 268.55 (0.74)& 256.80 (0.72)& 250.87 (0.72)& 246.49 (0.73)& 252.02 (0.74)
\end{tabular}
\vspace{.5em}
\caption{Group-wise expected second stage sample size and overall power for different conditional error functions ($\delta_0 = 0.125$)}
\label{expectedsecondstagesamplesize}
\end{table}

We might not be satisfied with the achieved overall power at the pre-specified alternative $\delta =0.2$. In the following, we will show how to control the overall power by choice of the first stage sample size.

\subsection{Study design with overall power control}

The overall power of at least $0.8$ at the alternative $\delta = 0.2$ can be obtained by choice of the first stage sample size where the necessary first stage sample size can be determined using a root finding algorithm. In case of the inverse normal method with equally weighted stages ($w_1=w_2= \sqrt{1/2}$) we need a first stage sample size of $121$, for the inverse normal method with the weights $w_1=\sqrt{1/3}$, $w_2=\sqrt{2/3}$ we need  $193$ and with the optimal conditional error function for $\Delta = 0.2$ and the maximum likelihood approach the first stage sample size becomes $144$.

Given the mentioned first stage sample sizes, the maximum overall sample size is again largest for the inverse normal method with equally weighted stages ($1393$) and smaller for the inverse normal method with unequal weights $w_1=\sqrt{1/3}$, $w_2=\sqrt{2/3}$ ($1203$). For the optimal conditional error function at $\Delta = 0.2$ it is $1148$, and it is again smallest with the maximum likelihood approach ($831$). In table \ref{expectedoverallsamplesize} we see the expected overall sample size for a true $\delta$ of $0$, $0.125$ and $0.2$.

\begin{table}%[H]
\centering
\renewcommand{\arraystretch}{1.5}
\begin{tabular}{c|cccc}
$\delta$ &
 \parbox{3cm}{\centering Inv. normal \\($w_1=w_2=\sqrt{1/2}$)} & 
 \parbox{4cm}{\centering Inv. normal \\($w_1= \sqrt{1/3}$, $w_2=\sqrt{2/3}$)} &
 \parbox{3cm}{\centering Optimal monotone \\($\Delta = 0.2$)}&
 \parbox{3cm}{\centering Optimal monotone \\(Max. likel.)}\\ 
  \hline
0 & 475.26 & 523.74 & 443.87 & 411.71 \\ 
0.125 & 500.09 & 552.62 & 478.20 & 471.16 \\ 
0.2 & 393.07 & 414.97 & 383.36 & 393.18
\end{tabular}
\vspace{.5em}
\caption{Expected overall sample size for different conditional error functions}
\label{expectedoverallsamplesize}
\end{table}
Although the expected overall sample size is nearly the same for the equally weighted inverse normal method as for the optimal monotone conditional error function that uses the maximum likelihood approach, we save overall maximum sample size when using the optimal conditional error functions.

\section{Optimization under bounds for the sample size and CEF}\label{sec:bounds}
For financial and ethical reasons there is often a maximum sample size that cannot be exceeded, and there may be also a minimum for the second stage sample size, e.g.\ for statistical reasons (to be close enough to asymptotic limits) or scientific reasons, e.g.\ to collecting sufficient safety data. We show next how we can obtain the optimal monotone CEF under such additional constraints. The solution is simple and similar to the solution for the generally non-monotone optimal CEF considered in \cite{Brannath04}. Let $n_{2,min} < n_{2,max}$ be the desired minimum and maximum second stage sample sizes. A simple calculation shows that 
\begin{align*}
    n_{2,min} \leq d \cdot \bigr[\Phi^{-1}\{1-A(p_1)\}+\Phi^{-1}(cp)\bigl]^2/\tilde{\delta}(p_1)^2\leq n_{2,max}.
\end{align*}
is equivalent to the following constraint on the conditional error function
\begin{align}\label{eq:AConstraint}
    \alpha_{2,min}^n(p_1) \leq A(p_1)\leq \alpha_{2,max}^n(p_1),
\end{align}
for
\begin{align}\nonumber
    \alpha_{2,min}^n(p_1) := \Phi \bigr\{\Phi^{-1}(cp)-\tilde{\delta}(p_1)\sqrt{n_{2,max}/d}\bigl\} \quad\text{and}\hspace{3em}\\ 
    %\quad
    \alpha_{2,max}^n(p_1) := \Phi \bigr\{\Phi^{-1}(cp)-\tilde{\delta}(p_1)\sqrt{n_{2,min}/d}\bigl\}.
    \label{eq:Amima}
\end{align}

More generally, one can consider arbitrary constraint functions $\alpha_{2,min}(p_1)$ and $\alpha_{2,max}(p_1)$ that need to satisfy the properties
\begin{align}
\alpha_{2,min}(p_1) \leq \alpha_{2,max}(p_1) \leq cp \text{ for all } \alpha_1 < p_1 < \alpha_0
\label{cond1thmconstr}
\end{align} and 
\begin{align}
    \alpha_1 + \int_{\alpha_1}^{\alpha_0} \alpha_{2, min}(p_1)dp_1 \leq \alpha \quad\text{as well as}\quad \alpha_1 + \int_{\alpha_1}^{\alpha_0} \alpha_{2, max}(p_1)dp_1 \geq \alpha.
    \label{cond2thmconstr}
\end{align}
To ensure that the resulting optimal CEF is non-increasing, we additionally need to require that $\alpha_{2,min}(p_1)$ and $\alpha_{2,max}(p_1)$ are non-increasing. For $1-\Phi(2) \leq cp \leq \Phi(2)$ the optimal monotone conditional error function under the constraints (C1'), (C2), (C3) and the additional constraint 
\begin{align*}
(C4) \quad \alpha_{2,min}(p_1) \leq A(p_1)\leq \alpha_{2,max}(p_1).
\end{align*} is then given by
\begin{align}\label{acopt}
    \tilde{A}^C_{opt, c'}(p_1)= \max \bigl[\alpha_{2,min}(p_1), \min\bigr[\alpha_{2,max}(p_1), \psi\{-e^{c'}/\tilde{Q}(p_1)\}\bigl]\bigr].
\end{align}
For the general result, we refer to Theorem \ref{constr} in the appendix.

We illustrate now, how \eqref{acopt}, or more generally Theorem \ref{constr}, can be applied to account for restrictions on the minimum and maximum second stage sample size. With \eqref{eq:Amima}, the conditions \eqref{cond1thmconstr} are always fulfilled. However, when using a decreasing data dependent relative effect the functions $\alpha_{2,min}^n(p_1)$ and $\alpha_{2,max}^n(p_1)$ are not necessarily non-increasing and then the constrained optimal conditional error function may non-longer be non-increasing as well. However, the constraints $\alpha_{2,min}^n(p_1)$ and $\alpha_{2,max}^n(p_1)$ can instead be expressed with constant constraints on the conditional error function. In detail, the maximum resp. minimum second stage sample size is reached at $\alpha_0$ resp. $\alpha_1$ and these values can be plugged into \eqref{eq:Amima} to obtain the corresponding constant minimum resp. maximum conditional error.
% Theoretically, the issue can be resolved by replacing the non-monotonous constraints with upper and lower (non-increasing) envelopes for the lower and upper limits \eqref{eq:Amima}, respectively. Note that the envelopes also need to satisfy the conditions \eqref{cond2thmconstr}. Since such envelopes do not always exist or are not easy to determine, this approach has not been implemented in the \texttt{R} package \texttt{optconerrf}. In practice it is recommended to use non-increasing constraints on the conditional error function like e.g.\ a constant maximum and/or minimum for the conditional error function itself.

\subsection{Example: Design with constraints on the sample size}
\begin{figure}%[H]
\includegraphics[width=.8\linewidth]{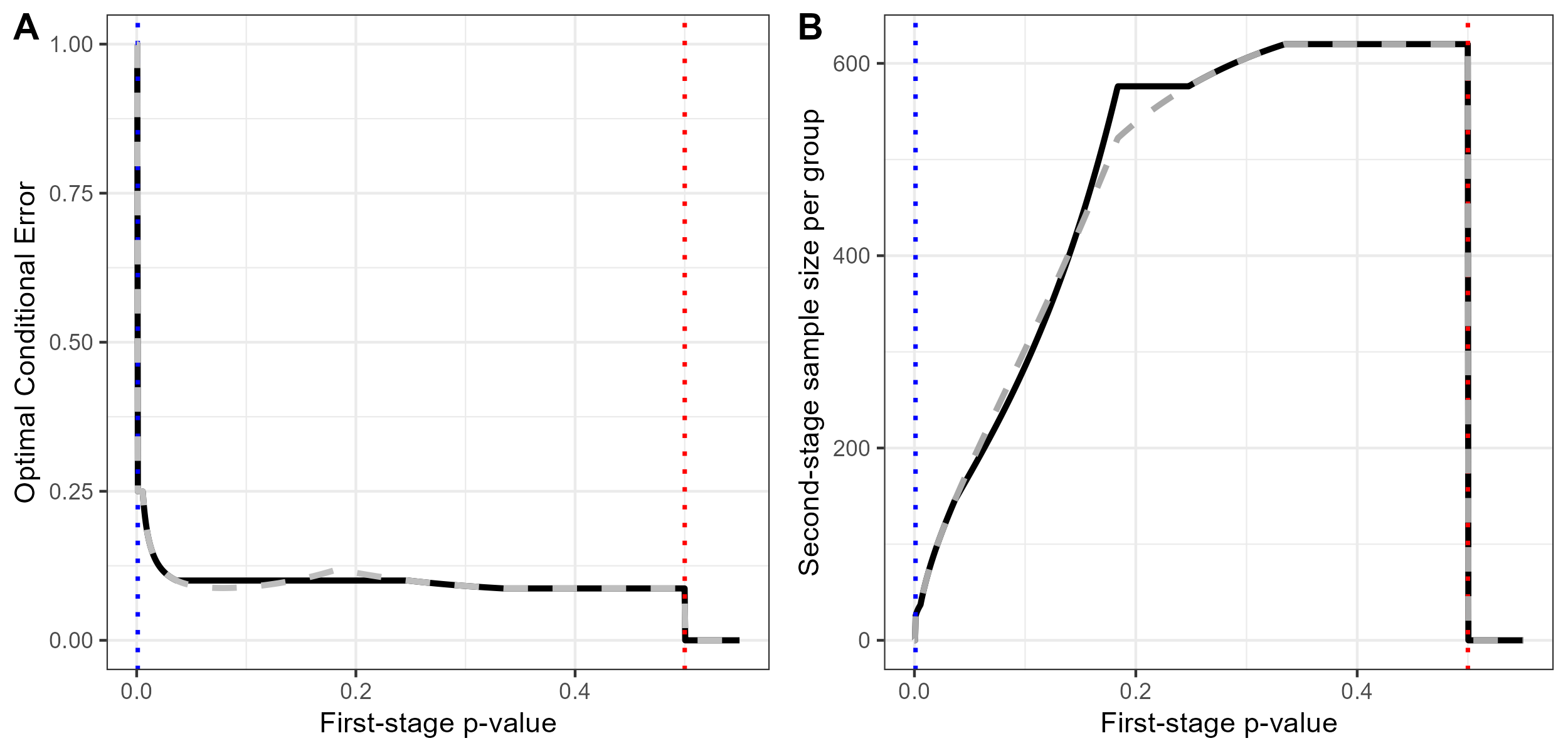}
\centering
\caption{\textbf{A:} Optimal CEF (dashed line) and optimal monotone CEF (solid line) with the constraints $n_{2,max}=620$ and $\alpha_{2,max}(p_1)= 0.25$ for the values described in the text, \textbf{B:} Corresponding second stage sample sizes}
\label{conderfun_constr}
\end{figure}

We consider now the example in Section \ref{sec:example} with the additional constraints $\alpha_{2, max}(p_1) = 0.25$ (not to test $H_0$ at a too large conditional level at stage 2) and $n_{2, max} = 620$. We determine the accordingly constrained optimal  monotone CEF based on the maximum likelihood approach, again using our \texttt{R}-package \texttt{optconerrf}. Figure \ref{conderfun_constr} \textbf{A} shows that the first constraint was met and in figure \ref{conderfun_constr} \textbf{B} we can see that the constraint automatically leads to a minimal second stage sample size. Furthermore, figure \ref{conderfun_constr} \textbf{B} shows that all p-values above a certain threshold lead to the maximum second stage sample size of $620$.

\subsection{Design example with an almost free lunch internal pilot study}

In this section we compare two potential designs with a data dependent sample size calculation. The first design (S) represents the standard approach, that consists of a separate pilot and confirmatory study, where the pilot study is only used for the calculation of the sample size of the  confirmatory part and not for testing the primary hypothesis $H_0$. The second design (A) is an adaptive two-stage design, where the pilot and confirmatory studies are integral parts of a single trial, corresponding to the first and second stage of the design, and the data from both stages are used to test $H_0$. We assume that in both designs,  the pilot study comprises $104$ subjects per group, and the sample size calculation is based on the pilot effect estimate, truncated from below by the minimal clinically relevant effect $\delta_0 = 0.125$. In both designs, we aim for a conditional power of $0.8$, whereby in design S this is just the unconditional power of the extra confirmatory trial. In design S there is no binding futility stopping foreseen, whereas in design A we stop at the pilot study for futility when $p_1> \alpha_0 = 0.9$.
This binding futility stopping rule is rather liberal and typically leads to an only small probability of stopping for futility under the alternative, e.g.\ only $0.003$ when the true effect is $\delta=0.2$.
No early rejection is foreseen, neither in design S nor in the adaptive design A ($\alpha_1 = 0$). In both designs, $H_0$ is tested at the nominal overall level $\alpha=0.05$. 

\begin{figure}%[H]
\includegraphics[width=.8\linewidth]{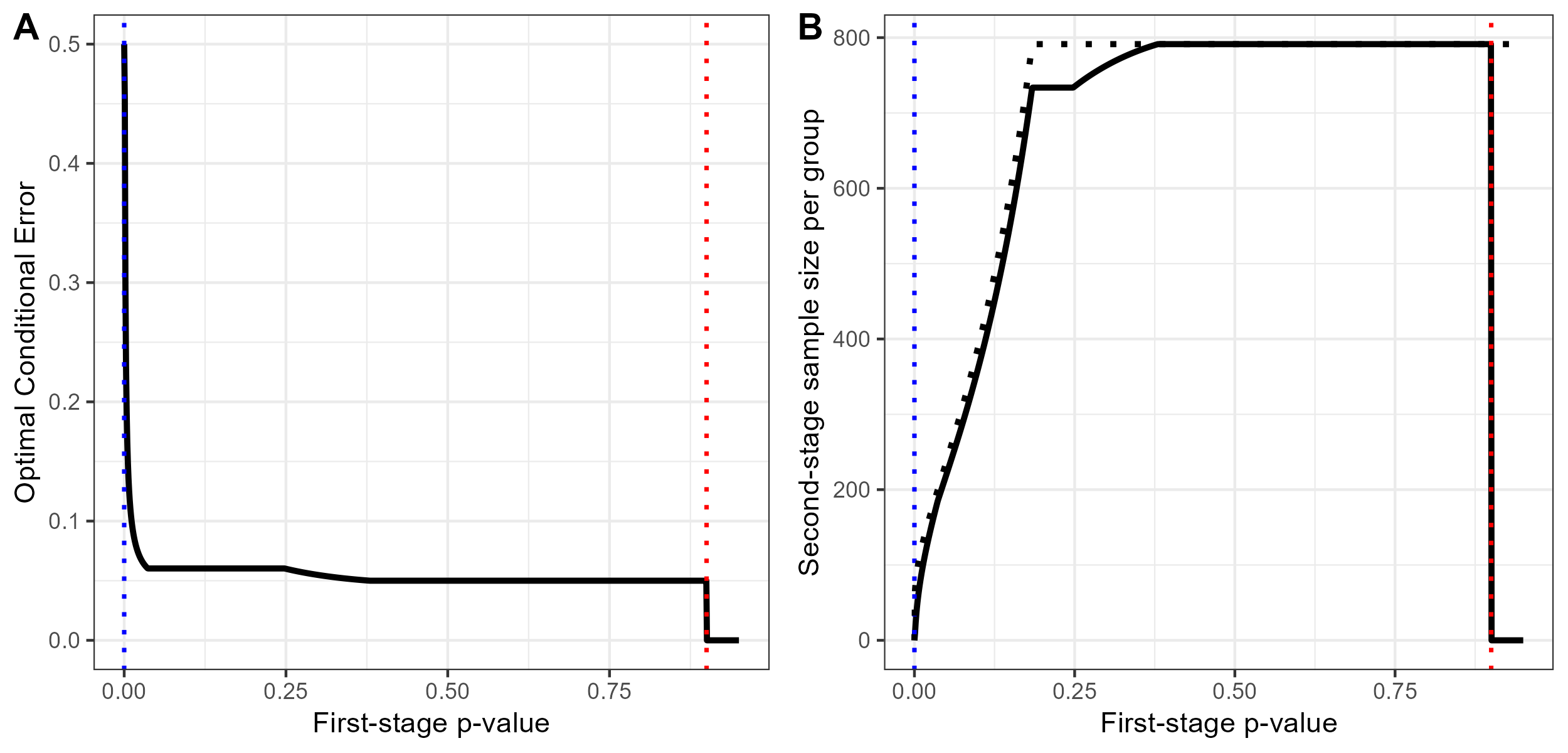}
\centering
\caption{\textbf{A:} CEF for the optimal adaptive design A \textbf{B:} Second stage sample size of the optimal adaptive design A (solid line) and the separate study design S (dotted line)}
\label{almost_free_luch}
\end{figure}

For the adaptive design we will use the optimal monotone conditional error function that uses the maximum likelihood approach. In addition, we apply the constant constraints $\alpha_{2, min} = 0.05$ and $\alpha_{2,max} = 0.5$ to the conditional error function. This means that with design A the significance level for the second stage is never smaller than the level $\alpha=0.05$ applied in design S to the confirmatory trial, and never exceeds $0.5$. Figure \ref{almost_free_luch} shows the optimal monotone conditional error function and the corresponding second stage sample size.
As expected, both designs have the same second stage maximum sample size ($792$). However, for small first stage p-values, the adaptive design A leads to sample sizes that are  smaller than those of design S. 

Moreover, one can see from Table \ref{freelunchexpectedsamplesize}, that design A also leads to smaller expected sample sizes and to higher overall power than design S. Since there is neither a loss in power nor in sample size, with the minor price of a (rather liberal) binding futility rule, we can think of receiving an almost free lunch with the adaptive design.

\begin{table}%[H]
\centering
\renewcommand{\arraystretch}{1.5}
\begin{tabular}{c|cc}
$\delta$ &
 \parbox{4cm}{\centering Separate study design S} & 
 \parbox{4cm}{\centering Optimal adaptive designs A}\\
  \hline
0 & 717.46 (0.05) & 625.52 (0.05)\\
0.125 & 549.65 (0.62) & 513.95 (0.63)\\ 
0.2 & 410.53 (0.74) & 378.46 (0.78)
\end{tabular}
\vspace{.5em}
\caption{Expected second stage sample size and overall power for the separate study design S and optimal adaptive design A}
\label{freelunchexpectedsamplesize}
\end{table}

\section{Summary and discussion}

In this work we extend the theory for optimal adaptive designs initiated in \cite{Brannath04} by a mathematical method that always provides the optimal non-increasing conditional error function and overcomes artificial constraints on the conditional power. Some of these extensions have already been indicated in \cite{Brannath04}, but have been worked-out, further extended and implemented only for this manuscript. We show that non-increasing conditional error functions are not only reasonable but also necessary for general type I error rate control with strictly conservative p-values, which are quite possible with one-sided null hypotheses. Utilizing the \texttt{R} package \texttt{optconerrf}, written to facilitate the application of the theoretical results presented in this manuscript, we illustrate by examples, that the desired monotonicity property is not always guaranteed with the original methodology suggested in \cite{Brannath04}. To obtain the optimal non-increasing CEF, the monotonicity can be enforced by a numerical procedure that monotonises the core function $Q$ of the (original) non-monotone optimal CEF. This approach preserves the flexibility of the original method, which permits to determine the minimal expected sample size (and corresponding optimal conditional error function) for any parameter value, as well as stochastic mixtures of parameter values represented by prior densities for the efficacy parameter. It also preserves the simple maximum likelihood approach of \cite{Brannath04}, that does not require the specification of a prior distribution and provides expected sample sizes that are close to the theoretical optima over a wide range of parameter values.  We have also seen that optimal conditional error functions do not only lead to smaller expected sample sizes but also reduce the maximum sample size, often more substantially than the expected one, in particular when using a prior or the maximum likelihood approach.  

We have extended the method to also account for constraints on the sample size and/or lower and upper bounds for the conditional error function. We have applied such a constrained and monotone optimal conditional error function to derive designs with an interim sample size calculation that uniformly improve the canonical approach of separate pilot and confirmatory study with respect to the sample size, and were seen to also lead to higher power for a range of typical parameter values. The advantage of adaptive designs over the separate study approach is particularly important for two-step registration processes, like the DiGA fast-track program of the German Federal Institute for Drugs and Medical Devices (BfArM), and has been investigated and illustrated in \cite{Kluge2025} with more classical approaches like the inverse normal method and Fisher's product test.        

The optimal non-increasing conditional error functions are constant on the non-monotonicity intervals of the original optimal CEF. We have also seen in examples that the original optimal CEF is already non-increasing (and hence no monotonising step is required) when truncating the interim estimate (used for the sample size determination) at a sufficiently high value. Since a constant conditional error function implies an only limited use of the first stage data, this indicates that a too extensive  use of the interim data (by e.g.\ truncating the interim estimate at an only small $\delta_0$) comes with too high costs in terms of sample size in order to be optimal. In practice one may therefore prefer a design where the monotonisation for the optimal conditional error function is not required. However, the theory presented in this paper provides a solution also for cases where this cannot be avoided due to clinical constraints.

It might be of interest to let the target conditional power to depend on the first stage evidence. The available resources for the second stage may increase with increasing first stage evidence, permitting a higher target conditional power for smaller p-values. Therefore, we could assume a given conditional power function $cp(p_1)$ that is non-increasing in the first stage p-value $p_1$. In this case it is still possible to use the described approach to derive a monotone conditional error function but this function is not necessarily optimal anymore. However, we assume that the derived function is still sufficiently efficient.
% So far we have reassessed the sample size to achieve a constant conditional power. However, there might be several reasons to let the target conditional power to depend on the first stage evidence. The available resources for the second stage may increase with increasing first stage evidence, permitting a higher target conditional power for smaller p-values. Moreover, in fixed size sample tests the conditional power typically increases with increasing evidence, including efficient and the most powerful tests. Hence, we may expect a gain in efficiency when using a target conditional power that is larger for smaller first stage p-values. Therefore, we assume now a given conditional power function $cp(p_1)$ that is non-increasing in the first stage p-value $p_1$. Note that using a non-increasing conditional power function is in line with using early rejection and futility boundaries.

Several extensions of our work are of interest. The most interesting question is how to define and derive optimal adaptive designs with also treatment, subgroup and/or endpoint selection. Another interesting question is how to obtain optimal adaptive designs that formally minimize a loss function that is e.g. based on the average and maximum sample size and whether this leads to substantial difference in the optimal conditional error functions.

We finally acknowledge the numerical optimization procedure implemented in the \texttt{R} package \texttt{adoptr}, which determines optimal adaptive designs by a purely numerically optimization procedure (after a discretisation of the sample and design space); see e.g.\ \cite{Kunzmann2021} and \cite{Pilz2024}. The theoretical approach presented here has the advantage of disclosing the mathematical structure of the optimal conditional error function and to providing a transparent, (almost) analytical solution, that is also easy to determine numerically. This does not only support the application of optimal adaptive designs in real clinical trials, but also provides a benchmark for more pragmatic approaches.

%\newpage
\bibliographystyle{unsrturl} 
\bibliography{references}

\newpage
%\appendix
\section*{Appendix}
%\newcounter{mycounter}
\renewcommand{\thesection}{\Alph{section}}
\setcounter{section}{1}
%\Alph{section}
\begin{proof}[Proof of Theorem \ref{nocpconstr}]
According to \cite{Wassmer2025} (chapter 8, p. 265)
minimizing the expected second stage sample size is achieved by maximizing for an arbitrary $c$ (to be chosen thereafter, to meet the level condition)
\begin{align*}
    \mathcal{G}_c(A) &= \int_{\alpha_1}^{\alpha_0}\int_{A(p_1)}^{cp} \{\nu'(u)Q(p_1)+e^c\}du\, dp_1\\
    &= \int_{\alpha_1}^{\alpha_0} Q(p_1)\int_{A(p_1)}^{cp} \{\nu'(u)+\frac{e^c}{Q(p_1)}\}du\, dp_1.
\end{align*}
For notational simplicity we define $x := -e^c/Q(p_1)$. Obviously, $\mathcal{G}_c(A)$ is maximal if we choose $A(p_1)=a$ such that
\begin{align}\label{maxint}
    \int_{a}^{cp} \big\{\nu'(u)-x\big\} \; du
\end{align}
is maximal. In order to maximize (\ref{maxint}), we need to know the monotonicity properties of the function 
\begin{align*}
    \nu'(u)&= -2 \{\Phi^{-1}(1-u)+k\}\sqrt{2\pi}e^{\{\Phi^{-1}(1-u)\}^2/2},\quad \text{where } k:= \Phi^{-1}(cp).
\end{align*}
Because the function $z(u)= \Phi^{-1}(1-u)$ is decreasing in $u$, the function $\nu'(u)$ is increasing in $u$, if the function 
\begin{align*}
    h(z)= (z+k)e^{z^2/2}
\end{align*}
is increasing in $z$. Therefore, we consider the derivative of the function $h(z)$
\begin{align}
    h'(z)&= e^{z^2/2}+(z+k)ze^{z^2/2} = e^{z^2/2} (z^2+zk+1) %\label{parabola}
    = e^{z^2/2}\{(z+\frac{k}{2})^2+1-\frac{k^2}{4}\} \label{ge0}.
\end{align}
From (\ref{ge0}) we see that $h'(z)>0$ for all $z$ if $1\geq k^2/4=\Phi^{-1}(cp)^2/4$. Hence, the function $\nu'(u)$ is increasing in $u$ for $1-\Phi(2) \le cp \le \Phi(2)$ and the integral (\ref{maxint}) can be maximized by setting $A(p_1)=\psi(x)$, where $\psi$ is the inverse of $\nu'(u)$.

We consider now the case $k^2/4>1$, i.e., where either $cp> \Phi(2)$ or $cp<1-\Phi(2)$. In this case, the derivative $h'(z)$ has the two roots
\begin{align*}
    z_{1/2}= -\frac{k}{2}\pm \sqrt{\frac{k^2}{4}-1},
\end{align*}
whereby $h'(z)<0$ for $z_1 < z < z_2$ and $h'(z)>0$ for $z < z_1$ or $z > z_2$. 
%because the function $z^2+zk+1$ describes a parabola that is opened upwards (\ref{parabola}). 
Defining 
\begin{align*}
    u_{1/2} := 1- \Phi \Bigl(- \frac{k}{2} \pm \sqrt{\frac{k^2}{4}-1}\Bigr),
\end{align*}
we obtain that $\nu'(u)$ is increasing in $u$ until $u_1$ then decreasing until $u_2$ and then increasing again. Consequently $\nu'(u_1)$ and $\nu'(u_2)$ are local maximum and minimum of $\nu'(u)$. The following arguments are best understood with a graphical sketch of the function $\nu'(u)$ (which we leave to the reader).

Let us return to the problem of maximizing the integral (\ref{maxint}). In case that $x<\nu'(u_2)$ respectively $x>\nu'(u_1)$, 
there exists at most one solution of the equation $\nu'(u)-x = 0$, which we denote by $\psi_l(p_1)$ respectively $\psi_u(p_1)$. 
Because $x=-e^c/Q(p_1)<0$ and $\lim \limits_{u \to 0} \nu'(u)= -\infty$, we always have a solution $\psi_l(p_1)\in \, ]0, u_1[$ if $x<\nu'(u_2)$; and because $\nu'(cp)=0$, we always  
have a solution $\psi_u(p_1)\in \, ]u_2, cp[$ if $x>\nu'(u_1)$. The solutions $\psi_l(x)$ and $\psi_u(x)$ can be found with a bisection search in the respective interval. Because the function $\nu'(u)-x$ is increasing for $u \in \,]0, u_1[$ and $u \in \,]u_2,cp[$, we know that $A(p_1) = \psi_l(x)$ respectively $A(p_1) = \psi_u(x)$ maximizes (\ref{maxint}).\\

The equality $\nu'(u)-x = 0$ has for $\nu'(u_2) < x < \nu'(u_1)$ three solutions and for $\nu'(u_2)=x$ or $\nu'(u_1)=x$ only two. Similar to the previous notation we denote the solution that lies in the interval $]0, u_1]$ with $\psi_l(p_1)$ and the solution that lies in the interval $[u_2, cp[$ with $\psi_u(p_1)$. The third solution, that only exists if $\nu'(u_2) < x < \nu'(u_1)$ and which we denote by $\psi_{mid}$,  lies between $u_1$ and $u_2$. Setting $A(p_1)$ to this third solution never maximizes the integral (\ref{maxint}) because $\nu'(u)-x$ is negative on $]\psi_{mid},u_2[$. To maximize the integral (\ref{maxint}) we should start integrating at $\psi_l:=\psi_l(x)$, if 
\begin{align}\label{eq:tpsint}
    \int_{\psi_l}^{\psi_u} \Bigl\{\nu'(u)-x\Bigr\} \, du > 0 
    \qquad\Leftrightarrow\qquad \frac{\nu(\psi_u)-\nu(\psi_l)}{\psi_u(p_1)-\psi_l(p_1)}&>x
\end{align}
and in all other cases at $\psi_u:=\psi_u(x)$. Overall we get for $cp <1-\Phi(2)$ or $cp >\Phi(2)$ the optimal conditional error function $A(p_1)= \tilde{\psi}\{-e^c/Q(p_1)\}$ with
\begin{align}\label{eq:tpsi}
    \tilde{\psi}(x)=
    \begin{cases}
    \psi_l & \text{for } x < \nu'(u_2) \text{ or } x < \frac{\nu(\psi_u)-\nu(\psi_l)}{\psi_u-\psi_l} \\
    \psi_u & \text{for } x > \nu'(u_1) \text{ or } x \geq \frac{\nu(\psi_u)-\nu(\psi_l)}{\psi_u-\psi_l} .
    \end{cases}
\end{align} 

Because $\psi_l(x)$ and $\psi_u(x)$ are both increasing in $x$, and the integral in \eqref{eq:tpsint} is decreasing in $x$, the function $\tilde{\psi}(x)$ is non-decreasing in $x$.
\end{proof}

\begin{proof}[Proof of Theorem \ref{inflation}]
Let $A(p_1)$ be a conditional error function that fulfills level condition (\ref{levelcond}) and is increasing in $p_1$ (almost surely) on an interval $]d_1, d_2[ \subseteq [\alpha_1,\alpha_0]$. We will construct a conservative first stage p-value that leads to a type I error inflation. 
To this end, we define a density for the first-stage p-value $p_1$ of the form
$$
f_c(p_1) = 
\begin{cases}
1 & \text{ for } p_1 \in ]0,d_1] \\
f_{cc}(p_1) & \text{ for } p_1 \in ]d_1,d_2] \\
1 & \text{ for }  p_1\in ]d_2,1],
\end{cases}
$$
where $f_{cc}(p_1)$ is a positive step-function with $\int_{d_1}^{d_2} f_{cc}(p_1) dp_1 = d_2-d_1$. To construct $f_{cc}$, note first that there exists a $d_m\in ]d_1,d_2[$ such that 
\begin{align}\label{eq:aineq1}
A(p_1) < \int_{d_1}^{d_2} A(u)du/(d_2-d_1) \quad\text{for almost all}\quad p_1\in ]d_1,d_m[
\end{align}
%and 
\begin{align}\label{eq:aineq2}
A(p_1) > \int_{d_1}^{d_2} A(u)du/(d_2-d_1) \quad\text{for almost all}\quad p_1\in ]d_m,d_2[.
\end{align}
This is because the (almost surely) increasing function $H(p_1):= A(p_1) - \int_{d_1}^{d_2} A(u)du/(d_2-d_1)$ satisfies $\int_{d_1}^{d_2} H(u)du=0$ and it's overall non-negativity or overall non-positivity on $]d_1,d_2[$ would imply $H(p_1)=0$ almost surely on $]d_1,d_2[$, which violates its monotonicity property. 

We define now $f_{cc}(p_1)$ for $p_1\in ]d_1,d_2[$ as
\begin{align*}
f_{cc}(p_1) := 
\begin{cases}
1-(d_2-d_m) & \text{ for } p_1 \in ]d_1,d_m[\\
1+(d_m-d_1) & \text{ for } p_1 \in [d_m,d_2[.
\end{cases}
\end{align*}
Now, from the inequalities \eqref{eq:aineq1} and \eqref{eq:aineq2} we obtain that
\begin{align*}
\int_{d_1}^{d_2} A(p_1) f_{cc}(p_1) dp_1 & = \int_{d_1}^{d_2} A(p_1) dp_1 - (d_2-d_m) \int_{d_1}^{d_m} A(p_1) dp_1 + (d_m-d_1) \int_{d_m}^{d_2} A(p_1) dp_1 \\
& > \int_{d_1}^{d_2} A(p_1) dp_1 - (d_2-d_m) (d_m-d_1) \frac{\int_{d_1}^{d_2} A(u) du}{d_2-d_1} + (d_m-d_1) (d_2-d_m) \frac{\int_{d_1}^{d_2} A(u) du}{d_2-d_1} \\
& = \int_{d_1}^{d_2} A(p_1) dp_1.
\end{align*}
Finally, this implies
\begin{align*}
\int_0^1 A(p_1) f_c(p_1) dp_1 & = \int_0^{d_1} A(p_1) dp_1 + \int_{d_1}^{d_2} A(p_1) f_{cc}(p_1) dp_1 + \int_{d_2}^1 A(p_1) dp_1 \\
                        & > \int_0^{d_1} A(p_1) dp_1 + \int_{d_1}^{d_2} A(p_1)  dp_1 + \int_{d_2}^1 A(p_1) dp_1= \int_0^1 A(p_1) dp_1=\alpha. 
\end{align*}
\end{proof}

\begin{proof}[Proof of Theorem \ref{typeIerror}]
Let $p_1$ be a conservative p-value with density function $f_c(x)$ and distribution function $F_c(x)$, i.e.\ $F_c(u)\leq u$ for all $u \in [0,1]$. Then there exists a random variable $U \sim U[0,1]$ such that $U \leq p_1$ almost surely. Because the conditional error function $A(p_1)$ is non-increasing, it holds $A(U)\geq A(p_1)$ almost surely, and it follows from level condition (\ref{levelcond}) that
\begin{align*}
    \alpha = \int_0^1 A(u) du =E[A(U)] \geq E[A(p_1)]=\int_0^1 A(p_1) f_c(p_1)dp_1.
\end{align*}
\end{proof}

To prove Theorem~\ref{opti}, we need the following lemma. 
\begin{lem}\label{optth2}
The function $\tilde{Q}(p_1):=\tilde{Q}^{(K)}_{q_K}(\cdot)$ is non-increasing on
$]\alpha_1,\alpha_0]$. Further,
for every non-decreasing and non-negative function $\eta(p_1)$
\begin{equation}\label{ieqac}
\int_{\alpha_1}^{\alpha_0}\,\eta(p_1)\,Q(p_1)\,d p_1\ge
\int_{\alpha_1}^{\alpha_0}\,\eta(p_1)\,\tilde{Q}(p_1)\,d p_1,
\end{equation}
and for every  measurable real function $\xi\{\cdot\}$
\begin{equation}\label{eqac}
\int_{\alpha_1}^{\alpha_0}\,\xi\{\tilde{Q}(p_1)\}\,Q(p_1)\,d p_1=
\int_{\alpha_1}^{\alpha_0}\,\xi\{\tilde{Q}(p_1)\}\,\tilde{Q}(p_1)\,d p_1.
\end{equation}
\end{lem}
%Lemma \ref{optth2} can be used to show theorem \ref{opti}.

\begin{proof}[Proof of Lemma \ref{optth2}]
All results are verified by induction in $k$, and for
convenience, we define $\tilde{Q}^{(0)}_{q_0}(\cdot)=Q(\cdot)$. 

We start
showing the monotonicity of $\tilde{Q}(p_1)$. Obviously, $Q(p_1)$
is non-increasing on $]\alpha_1,d_{l 1}]$, and by our assumptions, 
$Q(p_1)$ is non-increasing also on $]d_{u k},d_{l k+1}]$ for all $k=1,\ldots, K-1$. 
Hence, if $Q^{(k-1)}_{q_{k-1}}(p_1)$ is non-increasing on $]\alpha_1,d_{l k}]$ then $Q^{(k)}_{q_k}(p_1)$ is non-increasing
on $] \alpha_1,d_{l k+1}]$ by its construction.

We next prove (\ref{ieqac}) by showing
$\int_{\alpha_1}^{\alpha_0}\,\eta(p_1)\,\delta_k(p_1)\,d
p_1\ge 0$ with
$\delta_k(p_1)=\tilde{Q}^{(k-1)}_{q_{k-1}}(p_1)-\tilde{Q}^{(k)}_{q_k}(p_1)$
for all $k$ and every non-decreasing non-negative $\eta(p_1)$. By
definition~(\ref{tm}) we have $\delta_k(p_1)\le 0$ for $p_1\le
d_{l k}$ and $\delta_k(p_1)\ge 0$ for $p_1\ge d_{u k}$. Further,
$\delta_k(p_1)=Q(p_1)-q_k$ for $p_1\in D_k$, which is increasing.
Hence there is a number $d_{0 k}\in ]d_{l k},d_{u k}]$ such that
$\delta_k(p_1)\le 0$ for all $p_1\le d_{0 k}$ and
$\delta_k(p_1)\ge 0$ for all $p_1\ge d_{0 k}$. If $\eta(p_1)$ is
non-decreasing, then also $\eta(p_1)-\eta(d_{0 k})\le 0$ for
$p_1\le d_{0 k}$ and $\eta(p_1)-\eta(d_{0 k})\ge 0$ for $p_1\ge
d_{0 k}$, which implies $\{\eta(p_1)-\eta(d_{0
k})\}\cdot\delta_k(p_1)\ge 0$ for all $p_1\in
]\alpha_1,\alpha_0]$. Since $\delta_k(p_1)=0$ for $p_1\ge
d_{l k+1}$, and by the choice of $q_k$, we have
$\int_{\alpha_1}^{\alpha_0}\delta_k(p_1)\,d
p_1=0$, we obtain
$\int_{\alpha_1}^{\alpha_0}\,\eta(p_1)\,\delta_k(p_1)\,d
p_1 = \int_{\alpha_1}^{\alpha_0}\,\{\eta(p_1)-\eta(d_{0
k})\}\,\delta_k(p_1)\,d p_1\ge 0$.

To see (\ref{eqac}) notice that $\tilde{Q}^{(k)}(p_1)=q_k$ if $\delta_k(p_1)\not=0$. Hence,
for all $k$ and every measurable $\xi\{\cdot\}$ we get
$\int_{\alpha_1}^{\alpha_0}\,\xi\{\tilde{Q}^{(k)}(p_1)\}\,\delta_k(p_1)\,d p_1 = \xi\{q_k\}
\int_{\alpha_1}^{\alpha_0}\,\delta_k(p_1)\,d p_1=0$.\\
\end{proof}

\begin{proof}[Proof of Theorem \ref{opti}]
From Theorem~4.1 in \cite{Brannath04} we know that
$\int_{\alpha_1}^{\alpha_0}\,\nu\{A(p_1)\}\,\tilde{Q}(p_1)\,d p_1$
is uniquely minimized by the conditional error function $\tilde{A}_{opt,\, c_\alpha}(p_1)=\psi\{- e^{c_\alpha} / \tilde{Q}(p_1)\}$,
which is non-increasing by the monotonicity of $\psi(\cdot)$ and the monotonicity of $\tilde{Q}(p_1)$ (see Lemma \ref{optth2}).
Hence, if $A(p_1)$ is another non-increasing conditional error function which satisfies level condition (\ref{levelcond}), then by (\ref{ieqac}) and (\ref{eqac}):
$\int_{\alpha_1}^{\alpha_0}\,\nu\{A(p_1)\}\,Q(p_1)\,d p_1
\ge
\int_{\alpha_1}^{\alpha_0}\,\nu\{A(p_1)\}\,\tilde{Q}(p_1)\,d p_1
>
\int_{\alpha_1}^{\alpha_0}\,\nu\{\tilde{A}_{opt,c_\alpha}(p_1)\}\,\tilde{Q}(p_1)\,d p_1
 = \int_{\alpha_1}^{\alpha_0}\,\nu\{\tilde{A}_{opt,c_\alpha}(p_1)\}\,Q(p_1)\,d p_1.$
\\[-1em]
\end{proof}

\begin{thm} \label{constr}
Let $\alpha_{2,min}(p_1)$ and $\alpha_{2,max}(p_1)$ be non-increasing functions that satisfy the properties
\eqref{cond1thmconstr} and \eqref{cond2thmconstr}.
Let
$\tilde{Q}(p_1)$ be the monotonized version of $Q(p_1)$. 

For $1-\Phi(2)\le cp \le \Phi(2)$ let further 
$$\psi_c(x,p_1):=\max \bigl[\alpha_{2,min}(p_1), \min\bigr\{\alpha_{2,max}(p_1), \psi(x)\bigl\}\bigr]$$
and for $cp<1-\Phi(2)$ or $cp > \Phi(2)$ let 
\begin{align}\label{psicul}
\psi_{c,l}(x,p_1):=\max \bigl[\alpha_{2,min}(p_1), \min\bigr\{\alpha_{2,max}(p_1), \psi_l(x)\bigl\}\bigr]\text{ and }\\
\psi_{c,u}(x,p_1):=\max \bigl[\alpha_{2,min}(p_1), \min\bigr\{\alpha_{2,max}(p_1), \psi_u(x)\bigl\}\bigr]
\end{align}
where $\psi_l(x)$ is the solution of the equation $\nu'(u) = x $ in the interval $]0, u_1]$ and 
$\psi_u(x)$ is the solution of the equation $\nu'(u) = x$ in the interval $[u_2, cp[$.

The optimal monotone conditional error function under the constraints (C1'), (C2), (C3) and the additional constraint 
\begin{align*}
    (C4) \quad \alpha_{2,min}(p_1) \leq A(p_1)\leq \alpha_{2,max}(p_1).
\end{align*}
equals for $1-\Phi(2)\le cp\le \Phi(2)$ the expression
\begin{align}%\label{eq:ACopt}
    \tilde{A}^C_{opt, c'}(p_1):= \psi_c\bigr\{-e^{c'}/\tilde{Q}(p_1),p_1\bigl\}
\end{align}
and for $cp<1-\Phi(2)$ or $cp> \Phi(2)$ the expression 
\begin{align}\label{eq:ACopt}
    \tilde{A}^C_{opt, c'}(p_1)= \tilde{\psi}_c\bigr\{-e^{c'}/\tilde{Q}(p_1),p_1\bigl\},
\end{align}
where 
\begin{align}
    \tilde{\psi}_c(x,p_1)=
    \begin{cases}
    \psi_{c,l}(x,p_1) &\text{for } x < \nu'(u_2) \text{ or } x < \frac{\nu\big\{\psi_{c,u}(x,p_1)\big\}-\nu\big\{\psi_{c,l}(x,p_1)\big\}}{\psi_{c,u}(x,p_1)-\psi_{c,l}(x,p_1)} \\[.1em]
    \psi_{c,u}(x,p_1) &\text{for } x> \nu'(u_1) \text{ or } x \geq \frac{\nu\big\{\psi_{c,u}(x,p_1)\big\}-\nu\big\{\psi_{c,l}(x,p_1)\big\}}{\psi_{c,u}(x,p_1)-\psi_{c,l}(x,p_1)}.
    \end{cases}
\end{align}
\end{thm}
\begin{proof}[Proof of Theorem \ref{constr}]
According to the proof of Theorem~\ref{nocpconstr}, for any given $p_1$, we need now to search for $a\in [\alpha_{2,min}(p_1),\alpha_{2,max}(p_1)]$ such that \eqref{maxint} is maximized. If $\alpha_{2,max}(p_1)$ is below the root $\psi(x,p_1)$ or the two roots $\psi_{l}(x,p_1)$ and $\psi_{u}(x,p_1)$, then $\nu'(u)-x$ is negative for all $u\in ]\alpha_{2,min}(p_1),\alpha_{2,max}(p_1)[$ and \eqref{maxint} is maximized for $\tilde{A}^C_{opt, c'}(p_1)=\alpha_{2,max}(p_1)$. If  $\alpha_{2,min}(p_1)$ is above the single or the two roots, then  $\nu'(u)- x$ is positive for all $u\in ]\alpha_{2,min}(p_1),\alpha_{2,max}(p_1)[$ and \eqref{maxint} is maximized for $\tilde{A}^C_{opt, c'}(p_1)=\alpha_{2,min}(p_1)$. 
If $1-\Phi(2)\leq cp\leq \Phi(2)$ and $\psi_{u,c}(x,p_1)\in [\alpha_{2,min}(p_1),\alpha_{2,max}(p_1)]$, then  \eqref{maxint} is maximized for $\tilde{A}^C_{opt, c'}(p_1)=\psi_{u,c}(x,p_1)$. If $cp > \Phi(2)$ or $cp < 1-\Phi(2)$ then
the $a$ maximizing \eqref{maxint} depends on which of the two roots belong to $[\alpha_{2,min}(p_1),\alpha_{2,max}(p_1)]$. Replacing  $\psi_{l}(x,p_1)$ and $\psi_{u}(x,p_1)$ by $\psi_{l,c}(x,p_1)$ and  $\psi_{u,c}(x,p_1)$ defined in \eqref{psicul}, one can see as in the proof of Theorem~\ref{nocpconstr}, that the optimal conditional error function is given by \eqref{eq:ACopt}.

That $\tilde{A}^C_{opt, c'}(p_1)$ is non-increasing in $p_1$ follows from the monotonicity of $\psi_c(x,p_1)$,  $\psi_{l,c}(x, p_1)$, $\psi_{u,c}(x,p_1)$ in $p_1$.
\end{proof}

\end{document}